\documentstyle[11pt,amsfonts]{article}
\renewcommand{\theequation}{\arabic{equation}}
\newcommand{\be}{\begin{equation}}
\newcommand{\ee}{\end{equation}}
\newcommand{\bea}{\begin{array}}
\newcommand{\ea}{\end{array}}
\newcommand{\beqa}{\begin{eqnarray}}
\newcommand{\eeqa}{\end{eqnarray}}
\newcommand{\bean}{\begin{eqnarray*}}
\newcommand{\eean}{\end{eqnarray*}}

\def\up#1{\leavevmode \raise.16ex\hbox{#1}}

\newcommand{\gapproxeq}{\lower
 .7ex\hbox{$\;\stackrel{\textstyle >}{\sim}\;$}}
\newcommand{\lapproxeq}{\lower .7ex\hbox{$\;\stackrel
{\textstyle <}{\sim}\;$}}

\renewcommand{\theequation}{\thesection.\arabic{equation}}

\newcounter{appendice}
\newcommand{\appendice}
{
\setcounter{equation}{0}
\renewcommand{\theequation}{\Alph{appendice}.\arabic{equation}}
\addtocounter{appendice}{1}
{\bf Appendix \Alph{appendice}}
}

\def\thebibliography#1{{\bf REFERENCES\markboth
 {REFERENCES}{REFERENCES}}\list
 {[\arabic{enumi}]}{\settowidth\labelwidth{[#1]}\leftmargin\labelwidth
 \advance\leftmargin\labelsep
 \usecounter{enumi}}
 \def\newblock{\hskip .11em plus .33em minus -.07em}
 \sloppy
 \sfcode`\.=1000\relax}

\begin{document}

%\vspace*{5mm}

\centerline{ \LARGE  A New Class of Two-Dimensional Noncommutative Spaces} 

\vskip 2cm

\centerline{ {\sc A. Pinzul and A. Stern}  }

\vskip 1cm
\begin{center}
Department of Physics, University of Alabama,\\
Tuscaloosa, Alabama 35487, USA
\end{center}

\vskip 2cm

\vspace*{5mm}

\normalsize
\centerline{\bf ABSTRACT}
We find an infinite number
of noncommutative geometries  which posses a differential structure.
They generalize the two dimensional noncommutative plane, and have
infinite dimensional representations. 
Upon applying  generalized coherent states we are able to take
the continuum limit, where we recover the punctured plane with  non
constant Poisson structures.
\vskip 2cm
\vspace*{5mm}

\newpage
\scrollmode

\section{Introduction}
Although the noncommutative geometry program\cite{reviews} holds
promise for string theory, quantum gravity and renormalization
theory, one dilemma facing it is the
scarcity of suitable examples.  For the case of two dimensions, there are
essentially  only two such examples: the noncommutative plane (or torus) and the fuzzy sphere.
They are distinguished, in part, by the size of their
representations, the former being infinite and the latter finite.  The
main obstacle in
constructing more  such systems is the requirement that they posses
 a differential structure, a necessary first step to writing down noncommutative  field
 theories.\footnote{ On the
   other hand,  more possibilities exist if one doesn't make this demand.
   See for example \cite{corn}.} In this article, we find new examples
of noncommutative
geometries  which posses a differential structure, and whose continuum limits are two dimensional manifolds.  They are associated with  algebras of the form
\be [z,\bar z] = \Theta \label{zbz}\;,\ee where $\Theta$ is a function
of $z\bar z$.
 We find an infinite  number of  solutions consistent with this
 anzatz.  They include  the
noncommutative plane ($\Theta$=constant) as a special case.  As in that case, the
algebras have infinite dimensional representations. 

 The commutative
limit for these new examples is the punctured plane.  We deduce this
after applying generalized coherent states defined on the complex
plane.  A classical limit can be defined which preserves the
differential structure.   Upon taking it we
get a
singularity in the symplectic two form  at the origin.  At large
distances from the origin the Poisson structure approaches a constant.  Alternatively,
we can say that the full noncommutative theory provides a
regularization for such a singularity, and a possible interpretation of our
solutions is that they correspond to solitons on the noncommutative
plane, which go to point singularities in the commutative limit.
Analogous phenomena were found for the fuzzy sphere by
S. Vaidya.\cite{sv}  A  physical application of the solutions
found here could be in the description of vortices in the fractional 
quantum Hall
effect.  For this  we recall the recent proposal by Susskind for
writing  the fractional quantum Hall effect
in terms of noncommutative Chern-Simons theory.\cite{sus}, \cite{us2}  There, vortices were inserted
aposteriori on the noncommutative plane.  However it may be more natural to
start with our noncommutative spaces which already contain vortex-like
features.  We intend to report on this in
a future work.

 After a brief review of the exterior derivative for the noncommuting
 plane in sec. 2, we construct the new noncommutative spaces in
 sec. 3.
Two steps are needed in this construction.  One is to modify the
  procedure for taking the exterior derivative from that used on the
 noncommutative plane. The other is to insure that there is no trivializing map
 of this algebra, i.e. one that takes it to  the noncommutative plane.
Two distinct coherent state descriptions and star products are constructed for 
 the new noncommutative spaces in secs. 4 and 5. 
 The coherent states   in sec. 4 are the standard ones for the harmonic
 oscillator and the star product is a familiar one, known as
 the Voros star product\cite{Voros}.  The latter is a `equivalent' to the Moyal
 star product\cite{Moyal} and is thus applicable to the noncommuting
 plane.  When applying it to our noncommutative spaces, it, however,
 has the disadvantage that the commutative limit is obscure.
For this purpose we recall alternative procedures for constructing coherent states and their
 associated star products in sec. 5.  Particularly useful is one developed by us along
 with G. Alexanian in \cite{us}.  Although  the  star product in this section is
 more involved than the Voros star product, a closed integral
 expression for it nevertheless exists.  [Alternatively, a systematic derivative
 expansion can be given, and this is done in  Appendix A.]  The main
 advantage of this star product is that it is easy
 to take the commutative limit.  We do this in sec. 6.  There we show
 how the singularity
 develops  in the limit.  It can be interpreted as a coordinate
 singularity on a K\"ahler manifold ${\cal M}$.   A procedure for
 quantizing  K\"ahler manifolds was given a long time ago by
 Berezin.\cite{brzn}  We prove in  appendix B  that Berezin's
 quantization procedure is not the inverse of our  classical
 limiting procedure, and it appears not to preserve the differential
 calculus.  Therefore  our new
 noncommutative spaces could not have been discovered by applying
 Berezin quantization to  ${\cal M}$.

\section{ The  noncommutative plane }
\setcounter{equation}{0}

  The algebra ${\cal A}_{np}$ of the  noncommutative plane  is generated by the operator
$z$ and its  hermitian conjugate  $\bar z$, satisfying the commutation
relation
(\ref{zbz}) 
where $\Theta$ is a nonzero central element.  An
exterior derivative $d$ can be easily defined for any operator $A$ in ${\cal
  A}_{np}$ which is consistent with linearity and the Leibniz rule.
For this we introduce an operator $Q$, and write
\be d A=[Q,A]\label{daqa}\ee   For  $dz$ and
$d\bar z$ to be hermitian conjugates we need that $Q$ is
antihermitian, and  in order that $d^2=0$, the anticommutator $[Q,Q]_+$ should be in the
center of ${\cal A}_{np}\;.$  This is the case for
  \be Q=\Theta^{-1} z \bar\chi - \chi \bar
z \Theta^{-1}\label{odq}\ee where  $\chi$ and its hermitian conjugate
$\bar \chi$ are assumed to be Grassmann elements satisfying trivial anticommutation relations
\be  [\chi,\chi]_+= [\bar \chi,\chi]_+= [\bar\chi,\bar\chi]_+=0  \label{tacr}\ee
They are also assumed to commute with $z$ and $\bar z$.   As a result,
 $[Q,Q]_+= 2\Theta^{-1}  \chi\bar\chi$, which commutes with $z$ and
 $\bar z$,  and we can identify
 $\chi$ and
$\bar\chi$ with  $dz$ and
$d\bar z$, respectively.

\section{Constructing novel noncommutative  spaces}
\setcounter{equation}{0}

Now we generalize to algebras ${\cal A}_{new}$ where  $\Theta$ is a nonsingular
hermitian operator depending on $z$ and  $\bar z$.  We shall limit the
discussion to functions of $z\bar z$.   To define the
exterior derivative, we again assume the existence of an operator $Q$
such that (\ref{daqa}) holds.  Once again we need that  $Q$  is
antihermitian, and   that  the anticommutator $[Q,Q]_+$ should be in the
center of the algebra.  

As a preliminary step, let us 
 see what happens if we keep the  definition
 $Q$  in (\ref{odq}), with
  $\chi$ and 
$\bar \chi$ once again satisfying the trivial anticommutation relations
(\ref{tacr}).\footnote{ We thank    P. Pre\u snajder  for making this suggestion.}   Now we can no longer identify
 $\chi$ and
$\bar\chi$ with  $dz$ and
$d\bar z$, respectively, but rather
\beqa dz&=&[Q,z]\;=\;(1-\bar z[\Theta^{-1},z])\;\chi + [\Theta^{-1},z]z\;
\bar\chi    \cr & &\cr
 d\bar z&=&[Q,\bar z]\;=\; - \bar
 z[\Theta^{-1},\bar z]\;\chi +  (1+[\Theta^{-1},\bar z]z)\;\bar \chi \eeqa  
For    $[Q,Q]_+$ to be in the center of the  algebra the following
condition on $\Theta^{-1}$ must be satisfied:
\beqa [\bar z  \Theta^{-1},  \Theta^{-1} z] &=&
 \Theta^{-1} +  \Theta^{-1}\bar z z\Theta^{-1} -
\bar z  \Theta^{-2}z  \cr & &\cr &=&\Theta^{-1}_0 \;=\;
 {\rm central}\;{\rm element}\label{qdtz}\eeqa 
Note that $\bar z  \Theta^{-1}$ and $  \Theta^{-1} z$ generate the
 algebra of the noncommuting plane, so that any solution we find for
 ${\cal A}_{new}$ will have to contain  ${\cal A}_{np}$ as a subalgebra.

Next consider a mapping from  the noncommutative plane,  generated by
 raising and lowering operators ${\bf a}$ and ${\bf a}^\dagger$, with 
$[{\bf a},{\bf a}^\dagger]=1$.  We write the map as \footnote{ The same
 map appears in \cite{corn}.}
\be z=f({\bf n}+1) {\bf a} \;, \label{mza}\ee
where $f({\bf n}+1)$ is a hermitian function of the number operator  $ {\bf n}={\bf
  a}^\dagger{\bf  a}\;$.  We denote eigenvectors of the latter by
$|n>$, $n=0,1,2,...$, which  span the Hilbert
space  $H^{(0)}$, $\;{\bf n}|n>=n|n>$.   Then 
$\Theta(z\bar z)$ can also be expressed as a function of ${\bf n}$ 
\be\Theta(z\bar z)\equiv\tilde\Theta({\bf n})= ({\bf n}+1)f({\bf n}+1)^2-{\bf
  n}f({\bf n})^2 \label{titof}\ee  and consequently
\be f(n+1)^2 = \frac{ \Sigma(n)}{n+1}\;,\qquad
\Sigma(n)\equiv\sum_{m=0}^n \tilde\Theta(m) \label{fft}\ee  Now
examine the condition (\ref{qdtz}).  
   From its vacuum expectation value
 \be \tilde\Theta(0)^{-1} =\Theta^{-1}_0\;, \label{ivti}\ee
while  from
 expectation value for the first excited
 state we can write 
\be 
\tilde\Theta(1)^{-2}=2\;\Theta^{-1}_0\Sigma(1)^{-1}\;,\label{thon} \ee  where we used (\ref{fft}).
Repeated application of this procedure to the higher excited states gives
\be 
\tilde\Theta(n)^{-2}=(n+1)\;\Theta^{-1}_0\Sigma(n)^{-1} \label{ndneqfmo}\ee
 and also
\be \tilde\Theta(n+1)^{-1} +  \Sigma(n) \;\biggl(
 \tilde\Theta(n+1)^{-2} -\tilde\Theta(n)^{-2} \biggr)\; =\Theta^{-1}_0
\;, \label{difeq}\ee 
Upon solving for  $\tilde\Theta(n+1)^{-1}$,  we get the following  recursion relation
\beqa \tilde\Theta(n+1)^{-1} &=&\frac 1{2 \Sigma(n) }\biggl\{
-1\pm\sqrt
{1+4 \Sigma(n) \biggl( \Theta_0^{-1}+ \tilde\Theta(n)^{-2}\;
 \Sigma(n)\biggr)} \;\biggr\}
\cr & & \cr &  =&\frac {\Theta_0 \tilde\Theta(n)^{-2}}{2(n+1) }\biggl\{
-1\pm\sqrt
{1+4(n+1)(n+2)  \Theta_0^{-2} \tilde\Theta(n)^{2}}\biggr\}\label{trr}\eeqa

If we choose the plus sign in (\ref{trr}) and insert the initial
value (\ref{ivti}) at $n=0$ it is easy to check that  $ \tilde\Theta(n)^{-1}
=\Theta^{-1}_0\; {\it for\; all\;}n\ge 0$.  So then the only solution
is the trivial one, i.e. the  noncommutative plane!   If we choose
 the minus sign in (\ref{trr}),  $\tilde\Theta(n)$ goes rapidly  to
 zero for large $n$.  Starting with the assumption $-  \Theta_0
 \tilde\Theta(n)^{-1}>>n\;,$ (\ref{trr}) gives\footnote{ Note that in
   this case  $  \Theta_0$ and $
 \tilde\Theta(n)\;,n\ge 1$ have opposite sign.}
\be \frac {\tilde\Theta(n+1)^{-1}}{\tilde\Theta(n)^{-1}} \rightarrow 
\frac{ - \Theta_0  \tilde\Theta(n)^{-1}}n >> 1\;,\qquad{\rm as}\quad  n\rightarrow \infty \ee  
This means that $   \tilde\Theta(n)^{-1}$ grows faster than an
exponential, which validates the starting assumption.  Since $
\tilde\Theta(n)^{-1}$ appears in $Q$, differentiation becomes
ill-defined in the limit of large   $ n$.  This is unacceptable, so
our preliminary attempt at finding nontrivial solutions for
$\tilde\Theta(n)$ fails.\footnote{ There is yet  another possibility which
  we shall not consider here, and that is to choose different signs in
  (\ref{trr}) for different values of $n$.  To avoid the previously mentioned problems in the
  asymptotic region we should require the plus sign as $n\rightarrow
  \infty $.   On the other hand, with a minus sign as $n\rightarrow
  \infty $,   $\tilde\Theta(n)$ starting from a fixed value goes rapidly to
 zero, and this   may be one way of ending up with an `edge' in the
 commutative limit.\cite{us2}  In either case, with switches in sign we should expect to obtain
  singularities upon taking the continuum limit, and these are not unlike
  the ones we  obtain in sec. 6.}

A successful deformation of the noncommuting
plane requires a more involved modification than what was tried
above.  With this in mind, 
we  follow the two steps given below. 

\noindent ${\bf Step\; 1}$ is to modify the definition of $Q$ in
(\ref{odq}).  We replace $\Theta^{-1}$ in $Q$ by an
operator $M$ which is the inverse of $\Theta$ everywhere except at a
finite number  of states.  We first look at the case of only one such
state,
 namely $|0>$.  Then
 \be Q=M({\bf n}) z \bar\chi - \chi \bar
z M({\bf n})\;,\qquad  M({\bf n})=\tilde\Theta({\bf n})^{-1} -\rho_0 \;\Theta^{-1}_0 \; |0><0|\;,
\label{nodq}\ee 
and (\ref{qdtz}) is replaced by
 \be\tilde \Theta M^2 +  M\bar z zM -
\bar z  M^{2}z =\Theta^{-1}_0 =
 {\rm central}\;{\rm element}\label{mqdtz}\ee The vacuum
 expectation value now gives \be
M(0)^2=\Theta^{-1}_0 \tilde\Theta(0)^{-1}  \ee and consequently using (\ref{nodq})
\be     \tilde\Theta(0)^{-1}  =      M(0)+\rho_0\Theta^{-1}_0
 =\frac12 \;\Theta^{-1}_0(\;1+2\rho_0+\sqrt{1+  4\rho_0 }\;)\label{vvt} \ee
From now on, our sign choice is plus to avoid the problems encountered
 previously  in the
  asymptotic region with the opposite choice.  We recover  the previous
  initial value $\tilde\Theta(0)^{-1} =\Theta^{-1}_0$ when
 $\rho_0=0$, and consequently the noncommutative plane.  For general $\rho_0$,  $\tilde\Theta(0)^{-1}$ can have any value greater than or
 equal to $\Theta^{-1}_0/4$. 
Since $M(1)= \tilde\Theta(1)^{-1}$, (\ref{thon}) once again holds and we can  then solve for $ \tilde\Theta(1)^{-1}$ 
$$ \tilde\Theta(1)^{-1}   =\frac 12\; \tilde\Theta(0)^{-1}\; \biggl(\;
-1 +\sqrt
{1+ 8\;\Theta_0^{-1}  \tilde\Theta(0)}\; \biggr) $$
\be =\frac14 \;\Theta^{-1}_0(\;1+2\rho_0+\sqrt{1+  4\rho_0 }\;)\;
\biggl(-1+\sqrt{ 1+\frac 4{\rho_0^2}(1+2\rho_0 - \sqrt{1+4\rho_0}\;)} \biggr)\label{fes}\ee which as expected reduces to $\Theta^{-1}_0$ when
 $\tilde\Theta(0)^{-1} =\Theta^{-1}_0.$
Moreover, (\ref{ndneqfmo}-\ref{trr}) for $n\ge 1$ are also still
valid, allowing us to generate all $\tilde\Theta(n)$.  For all 
 $\tilde\Theta(0)^{-1} \ne\Theta^{-1}_0,$ we obtain nontrivial solutions.

 To understand the  behavior at large $n$ we can replace
(\ref{difeq}) by the first order differential equation
\be \tilde\Theta(n)^{-1} + \;(n+1)\Theta_0 \tilde\Theta(n)^2 \;\frac d{dn}
 \tilde\Theta(n)^{-2}\;  =\Theta^{-1}_0\;,
 \label{difinteq}\ee where we used (\ref{ndneqfmo}).
The asymptotic solution for large $n$ is
\be  \tilde\Theta(n)^{-1} \rightarrow \Theta^{-1}_0 + \frac{C}{\sqrt{n}}
\qquad{\rm as}\qquad n\rightarrow \infty \;,\label{tfln} \ee
where $C$ is a constant.   Thus $ \tilde\Theta(n)$ is well behaved for
 $ n\rightarrow \infty $.  It approaches a
 constant value, so we can say that  ${\cal A}_{new}$ approaches a
 noncommutative plane in this limit.

Even though in the above we have managed to obtain a non-constant
solution for
$\tilde\Theta(n)$, the corresponding algebra generated by $z$ and
$\bar z$ is {\it equivalent} to that of  the standard
noncommutative plane due to  the map (\ref{mza}) -  {\it provided it
  is invertible} - as  it   only corresponds to a change of variables.   On
the other hand, if we can insure that   the map (\ref{mza}) is non invertible, we have an algebra that
 cannot be mapped back  to the 
noncommutative plane.  So

\noindent ${\bf Step\;2}\quad $ is to  make the map ${\bf a},{\bf a}^\dagger \rightarrow z,\bar z $ non invertible.
  We can implement this by requiring that $f(n)$ has one or more zeros for some $n\ge 1$. 
(Note that the value of  $f(0)$ is arbitrary.)  More specifically, we
shall look at the case where $f(1)= f(2)=\cdot \cdot\cdot=f(n_0)=0\;.$   We  first examine
the case
$n_0=1$.  It follows from (\ref{fft}) that $\tilde\Theta(0)$ vanishes.  In (\ref{vvt}) this
corresponds to the limit $\rho_0\rightarrow \infty$. 
Substituting $\Sigma(1)=\tilde\Theta(1)$ in   (\ref{thon}), or taking
the limit  $\rho_0\rightarrow \infty$ in (\ref{fes}), gives $$\tilde\Theta(1) =
\frac12\; \Theta_0$$  Then once again  (\ref{trr}) can  be used to
generate all
$\tilde\Theta(n), \;n\ge2$,
and the behavior  for large $n$ is
given by (\ref{tfln}).
   We note in this case that $M(0)$  is
singular, and hence differentiation is ill-defined,
at $n=0$.  But  the state $|0>$ is
not present in the 
Hilbert space for $z$ and $\bar z$, which we call  $H^{(1)}$,  because
$z\;|1>=0$, i.e.  $|1>$ is the lowest
weight state in $H^{(1)}$.

Steps 1 and 2 can be repeated in a more general setting to obtain
more inequivalent algebras.  These algebras are labeled by integer
$n_0$, and are infinite in number.  Following step 1 we modify
 the definition of $Q$ to one where $M(n)$ differs from
$\tilde\Theta^{-1}(n)$, now  for more than one value of $n$.  So we   generalize
$M(n)$ in (\ref{nodq}) 
to \be   M({\bf n})=\tilde\Theta({\bf n})^{-1} - \;\Theta^{-1}_0 \;\sum_{m=0}^{n_0-1}\rho_{m}\; |m><m|
\ee for some positive integer $n_0$.  Following step 2 we then take all
$\tilde\Theta(n),\;n=0,1,2,...,n_0-1$ to zero (or equivalently, all
$\rho_n\rightarrow \infty$), producing a non invertible map
(\ref{mza}).
(\ref{ndneqfmo}) now generalizes to \be M(n)^{2}=(n+1)\;\Theta^{-1}_0\Sigma(n)^{-1} \ee
$\tilde\Theta(n_0)$ is then determined
by setting $n=n_0$ and using
$\Sigma(n_0)^{-1}=\tilde\Theta(n_0)^{-1}=M(n_0)$: 
\be \tilde\Theta(n_0) =
\frac {\Theta_0}{n_0+1}\; \label{icft}\ee Once again  (\ref{trr}) can  be used to
generate all
$\tilde\Theta(n), \;n\ge n_0+1$.  So for example
\be \tilde\Theta(n_0+1) =   \biggl( 1+  \sqrt{\frac{5n_0+9}{n_0+1}} \biggr)\;
 \frac {\Theta_0}{2(n_0+2)}\;\;,\label{nzpo}\ee
 and the behavior  for large $n$ is
given by (\ref{tfln}).
 Now we have that   $M(n),\;n=0,1,2,...,n_0-1$ are 
singular in the limit, and  differentiation becomes ill-defined for
all $0\le n\le n_0-1$.   But all the corresponding states $|n>,\;n=0,1,2,...n_0-1$
are  eliminated from the resulting Hilbert space, denoted by  $H^{(n_0)}$, since
$z\;|n_0>=0$, and now $|n_0>$ is the lowest
weight state in $H^{(n_0)}$.

\section{Coherent states and star product - type I}
\setcounter{equation}{0}

In this section and the next we look at two different coherent state descriptions
 for the above system and construct the corresponding star products.
 The different
coherent state descriptions in this section and the next have certain
 advantages and disadvantages.  The    coherent states
  of this section, which we denote by $\widetilde{|\alpha >}$, have the advantage
 of simplicity, and  they lead to a well known star product. 

 We define coherent states  $\widetilde{|\alpha >}\in H^{(n_0)} $  for
complex $\alpha$ 
\begin{enumerate}
\item to satisfy  a partition of unity
\be 1 = \int d\tilde\mu (\alpha,\bar \alpha )\;  \widetilde{| \alpha >} \widetilde{ < \alpha |}
\; \label{tcomp}\ee for some integration measure $ d\tilde\mu
(\alpha,\bar\alpha )$,   
\item to be of unit norm $\widetilde{<\alpha|}\widetilde{\alpha>}=1$ and 
\item to form  rays with respect to the action  of the two-dimensional
translation group generated by $ \bar z \Theta^{-1} $ and
$\Theta^{-1}z$.
\end{enumerate}

We first introduce the unitary operators
\be  U(\alpha,\bar\alpha) =e^{ \alpha \bar z \Theta^{-1} 
- \bar \alpha \Theta^{-1}z}  \ee  
From (\ref{qdtz})  we have the
composition rule \be U(\alpha+\beta,\bar\alpha+\bar\beta) = e^{-{1\over 2}\Theta_0^{-1} (\bar\alpha \beta -\bar\beta \alpha)} 
U(\beta,\bar\beta) U(\alpha,\bar\alpha) \label{protg}\ee
 and so the set  $\{U(\beta,\bar\beta)\}
$ forms the projective representation of the translation group.
We define the  coherent states according to \beqa \widetilde{|\alpha>} &=&U(\alpha,\bar\alpha)|n_0> \cr & &\cr
&=&e^{-\frac12|\alpha|^2 \Theta^{-1}_0}e^{\alpha \bar z
\Theta^{-1}} e^{- \bar \alpha \Theta^{-1}z}|n_0>\cr & &\cr
&=&e^{-\frac12|\alpha|^2 \Theta^{-1}_0}e^{\alpha \bar z
\Theta^{-1}} |n_0> \label{nsocs}\eeqa   From  $\{\widetilde{|\alpha
>}\}$ we thus have orbits in $H^{(n_0)}$ passing through the point $|n_0>$.
When acted on by $U(\beta,\bar\beta)$, these states transform as
$$U(\beta,\bar\beta) \widetilde{|\alpha>} =e^{{1\over{ 2}}\Theta^{-1}_0 (\bar\alpha \beta -\bar\beta \alpha)}
 \widetilde{|\alpha+\beta>}\;, $$ which follows from (\ref{protg}),
 and so they define rays under the
action of the translation group.  The coherent states  $\widetilde{|\alpha >}$
 diagonalize the operator 
$\Theta^{-1}z$.  From (\ref{nsocs})
\be \Theta_0\Theta^{-1}z\widetilde{|\alpha >}= \alpha
\widetilde{|\alpha >}\label{eeftz} \ee 
We can therefore identify the coordinate $\alpha$ with the expectation
value of  $ \Theta_0\Theta^{-1}z$:
 $$\alpha =\Theta_0\widetilde{<\alpha |} \Theta^{-1}z\widetilde{|\alpha >}$$

 Scalar products are easy to compute in
 this basis, and they are the essential ingredient for obtaining the
 measure, and also for  constructing the
 star product.  From (\ref{nsocs}) \beqa \widetilde{<\alpha}|\widetilde{\beta >} &=&<n_0|
U(\alpha,\bar\alpha)^\dagger U(\beta,\bar\beta) |n_0>=<n_0|
U(-\alpha,-\bar\alpha) U(\beta,\bar\beta) |n_0>\cr & &\cr&=&e^{{1\over{ 2}}\Theta^{-1}_0 (\bar\alpha \beta -\bar\beta \alpha)}<n_0|
U(\beta -\alpha,\bar\beta -\bar\alpha) |n_0>=e^{{1\over{
      2}}\Theta^{-1}_0 (2\bar\alpha \beta -|\alpha|^2 -|\beta
  |^2)}\eeqa
and hence \be   |\widetilde{<\alpha}|\widetilde{\beta >}|^2 =
e^{-\Theta^{-1}_0 |\alpha- \beta|^2} \ee
which is identical to the result for standard coherent states.  It
follows that the measure and also the star product are identical to the result for standard
coherent states.  The former
 is 
\be d\tilde\mu (\alpha,\bar\alpha )\;   =\frac 1\pi\Theta^{-1}_0  d\alpha_R \;d\alpha_I \;,
\label{imscs}\ee  and from it one gets the partition of unity. 
($\alpha_R$ and $\alpha_I $ denote the real and imaginary parts of $\alpha$.) 
The latter  is known as the Voros star product\cite{Voros} and it is
equivalent to the Moyal star product.\cite{Moyal}\cite{zac}  It is therefore relevant
for the  noncommutative plane.
Following \cite{brzn}, the  covariant symbol
 $\tilde{\cal
  A}(\alpha,\bar \alpha)$ of an operator $A$ on $H^{(n_0)}$ is defined by  $\tilde{\cal A}(\alpha,\bar \alpha
)=\widetilde{<\alpha |} A \widetilde{|\alpha
>}$.
The star product $\tilde\star$  between
any two covariant symbols $\tilde{\cal A}(\alpha,\bar \alpha
)$ and ${\cal B}(\alpha,\bar \alpha )$  associated with  operators $A$ and
$B$ is defined to be the covariant symbol of the product of operators:
 $$ [\tilde{\cal A}  \tilde\star \tilde{\cal B}](\alpha,\bar \alpha )
\;=\widetilde{<\alpha |} AB\widetilde{|\alpha >}\; $$ and here
$$\tilde\star=\exp{\;\Theta_0 \;\overleftarrow{ \frac\partial{ \partial\alpha }}\;\;
\overrightarrow{ {\partial\over{ \partial\bar\alpha} }}}$$

Another advantage of this basis  comes from the computation of
derivatives of covariant symbols.  Under infinitesimal translations parameterized
by $\epsilon$ and $\bar\epsilon$ \beqa\tilde{\cal A}(\alpha,\bar \alpha
)&\rightarrow &  \tilde{\cal A}(\alpha+\epsilon,\bar \alpha
+\bar\epsilon)=\widetilde{<\alpha |}U(\epsilon,\bar\epsilon)^\dagger A U(\epsilon,\bar\epsilon)  \widetilde{|\alpha
>} \cr & &\cr & &= \tilde{\cal A}(\alpha,\bar \alpha
)+\epsilon \widetilde{<\alpha |} [A,\bar z \Theta^{-1}] \widetilde{|\alpha
>}-\bar\epsilon \widetilde{<\alpha |} [A, \Theta^{-1}z] \widetilde{|\alpha
>}\eeqa  Therefore  \beqa \frac\partial{ \partial\alpha }\;\;
\tilde{\cal A}(\alpha,\bar \alpha
)&=&\widetilde{<\alpha |} [A,\bar z \Theta^{-1}] \widetilde{|\alpha
>}\cr
& &\cr  {\partial\over{ \partial\bar\alpha}}\;\;
\tilde{\cal A}(\alpha,\bar \alpha
)&=&- \widetilde{<\alpha |} [A, \Theta^{-1}z] \widetilde{|\alpha
>}\eeqa

In conclusion, by utilizing the operators  $ \bar z \Theta^{-1} $ and
$\Theta^{-1}z$ we have recovered the star product for the 
noncommuting plane.  This is not surprising since the operators  $ \bar z \Theta^{-1} $ and
$\Theta^{-1}z$ generate the noncommuting plane.  By taking
the limit $\Theta_0\rightarrow 0$ we then
recover the commutative plane parameterized by $\alpha$ and $\bar
\alpha$, attached with a constant
Poisson structure.  But we claim
that  this is  not the appropriate commutative limit of the
noncommutative theory generated by  
operators $z$ and $\bar z$.  The correct
commutative theory should be expressed in terms of coordinates
which are  the
commuting analogues of  $z$ and
$\bar z$, along with a Poisson bracket written in terms of these
coordinates.  The Poisson bracket  should be the classical 
analogue of the commutator, so we should get a non constant Poisson structure.    The appropriate coordinates are the covariant
symbols of $z$ and $\bar z$, which in our basis are  $\widetilde{<\alpha |} z \widetilde{|\alpha
>}$ and $\widetilde{<\alpha |} \bar z \widetilde{|\alpha
>}$.  From (\ref{eeftz}), $$\widetilde{<\alpha |} z \widetilde{|\alpha
>}=\alpha\Theta_0^{-1}\;\widetilde{<\alpha |} \Theta \widetilde{|\alpha
>}\;.$$  However,  we don't have an explicit expression for the covariant symbol
of $\Theta$ in terms of $\alpha$ and $\bar\alpha$, nor consequently
 $z$ and $\bar z$.  Furthermore, the result will be non analytic.
 These are  the main disadvantages of the coherent states
 $\{\widetilde{|\alpha >}\}$, as without knowing the appropriate
 change of variables  $\alpha,\bar\alpha \rightarrow \widetilde{<\alpha |} z \widetilde{|\alpha
>},\widetilde{<\alpha |} \bar z \widetilde{|\alpha
>}$,
 the commutative limit is obscure.

\section{Coherent states and star product - type II }
\setcounter{equation}{0}

A more appropriate basis of coherent states  for understanding the
 commutative limit would
 be one that diagonalizes $z$ (or $\bar z$), since
 then no change of variables is necessary.  
The general procedure for constructing such coherent states was given
 in \cite{mmsz},\cite{us}, and it will be applied to our algebra in this
 section.    We  denote  eigenvectors of $z$ by $|\zeta >$  for
complex $\zeta$.    Unlike with the coherent states of the previous section, the
 resulting star product does not have a simple form (except for the
 case  $\Theta=$constant,  where we once again get  the Voros
 star product).  Nevertheless, a closed
 integral expression can be given, which can also be evaluated order by 
order in a derivative expansion.[See appendix A.]
 
 Like the coherent states of the previous section, we require $|\zeta >$ to satisfy  1. the partition of
 unity [now with 
 some new
 integration measure $ d\mu
(\zeta,\bar\zeta )$], along with 2. the unit norm condition. 
 Here we replace condition 3. with the requirement that  $|\zeta >$ be an eigenstate of $z$
\be z|\zeta > = \zeta |\zeta >\; \label{zee}\ee  We now give the construction
for  $|\zeta >$.  Analogous to the states of the previous section,  we
 can write $|\zeta >$ (up to an overall normalization)    by acting
on the ground state $|n_0>$ with some operator,  though the  operator in this
 case will not be unitary.

From (\ref{fft}) we have
 $z\bar z = \Sigma({\bf n})$, and it follows that
$$ [\;z\;,\;\bar z \Sigma({\bf n})^{-1}]\;|\zeta > =( 1\;-\;\zeta \bar z
\Sigma({\bf n})^{-1})\;|\zeta > $$ and so
\beqa  z|\zeta >\; -\;  \frac 1{ 1\;-\;\zeta \bar z
\Sigma({\bf n})^{-1}}\;\;z\;\;  ( 1\;-\;\zeta \bar z
\Sigma({\bf n})^{-1})&|\zeta> & \cr &&\cr
=\; \biggl[z\;,\;  \frac 1{ 1\;-\;\zeta \bar z
\Sigma({\bf n})^{-1}} \biggr]\; ( 1\;-\;\zeta \bar z
\Sigma({\bf n})^{-1})&|\zeta >& \cr &&\cr =\;-\; \frac 1{ 1\;-\;\zeta \bar z
\Sigma({\bf n})^{-1}} \; \;[\;z\;,\; 1\;-\;\zeta \bar z
\Sigma({\bf n})^{-1}]& |\zeta >& \; =\; \zeta |\zeta > \eeqa
For this to be consistent with (\ref{zee}), $z$ should annihilate $ ( 1\;-\;\zeta \bar z
\Sigma({\bf n})^{-1})\;|\zeta> $.  The latter must therefore be
proportional to the lowest weight state $|n_0>$, and thus
\be ( 1\;-\;\zeta \bar z
\Sigma({\bf n})^{-1})\;|\zeta>= {\cal   N} (|\zeta |^2  )^{-\frac12}\;
|n_0 > \label{nnz} \ee or 
\be  {\cal N} (|\zeta |^2  )^{\frac12}\;
|\zeta > =  \frac 1{ 1\;-\;\zeta \bar z
\Sigma({\bf n})^{-1}}\;|n_0> \ee
 Upon Taylor expanding, 
this agrees with the expression for the deformed
coherent states of \cite{mmsz},\cite{us}  \be {\cal N} (|\zeta |^2  )^{\frac12}\;
|\zeta > = 
 \; |n_0>\; + \;\sum_{n=1}^{\infty}\;\frac{\zeta^n \;\;
|n+n_0>} { \;\sqrt{\Sigma(n_0)\Sigma(n_0+1)\cdot\cdot\cdot\Sigma(n_0+n-1)}}\label{ndcs}\ee
Requiring $|\zeta >$  to be of unit norm fixes ${\cal N} (|\zeta |^2  )  $,
\be {\cal N} (x  )  =1\;+\;\; \sum_{n=1}^{\infty}\;\biggl(\prod_{m=n_0}^{n-1+n_0}\Sigma(m) \biggr)^{-1}\; x^{n}
 \;. \label{tefn}\ee 
Alternatively, we can get another expression for ${\cal N}(x)$ by first
 taking the scalar product of (\ref{nnz}) with
 $|n_0>$ to get $<n_0|\zeta> = {\cal N} (|\zeta |^2  )^{-1/2}$,  and
 then with $|\zeta >$  to get
 \beqa
 {\cal N} (|\zeta |^2  )^{-1}&=&1\; -\;|\zeta |^2\;
 <\zeta |\Sigma({\bf n})^{-1} |\zeta >\cr & & \cr &=& 
 <\zeta |\;1-\bar z\Sigma({\bf n})^{-1}z\; |\zeta > \label{nas} \eeqa
${\cal N} (|\zeta |^2  )^{-1}$ is then said to be the covariant symbol of the operator
$\;1-\bar z\Sigma({\bf n})^{-1}z\;$.
Another useful relation is
\be
 \frac d{dx}\ln{{\cal N}(x)}|_{x=|\zeta |^2  }\;=
\; <\zeta |({\bf n}+1-n_0)\Sigma({\bf n})^{-1} |\zeta >\;, \label{ddxln} \ee
which can be easily verified from the Taylor expansion for ${\cal N}(x)^{-1}$
[c.f. (\ref{tefn})], and so $ \frac d{dx}\ln{{\cal N}(x)}$ is the covariant symbol of the operator
$({\bf n}+1-n_0)\Sigma({\bf n})^{-1}\;$.

Using the above coherent states, a new star product $\star$ can be
defined between any two covariant symbols ${\cal A}(\zeta,\bar \zeta )=<\zeta | A|\zeta
>$ and ${\cal B}(\zeta,\bar \zeta )=<\zeta | B|\zeta >$, where $A$ and
$B$ 
are operators acting on
$H^{(n_0)}$.  It is such that
 $$ [{\cal A}  \star {\cal B}](\zeta,\bar \zeta )
\;=<\zeta | AB|\zeta >\; $$  For standard
 coherent states 
where  $\tilde\Theta({\bf n})=\Theta_0$= constant, it is once again the  Voros star
product
$\star=\exp{\;\Theta_0 \;\overleftarrow{ \frac\partial{ \partial\zeta }}\;\;
\overrightarrow{ {\partial\over{ \partial\bar\zeta} }}}$.
More generally, an explicit expression for it using generalized or deformed coherent
states on the plane was given in \cite{us}, and it depends implicitly
on $ {\cal N} (x)$.  By performing a derivative expansion, which we do
in  appendix A, one obtains the following leading three terms acting on
functions of $\zeta$ and $\bar\zeta$:
\be\star\;\; =\;\;  1\;\;+ \;\;\overleftarrow{ \frac\partial{ \partial\zeta }}\; \theta(|\zeta|^2) \;
\overrightarrow{ {\partial\over{ \partial\bar\zeta} }}\;\;
+\;\;
\frac14\biggl[\overleftarrow{\frac{\partial^2}{ \partial\zeta^2 }}
    \;\;  \overrightarrow{ \frac\partial {
      \partial\bar\zeta } }\theta(|\zeta |^2)^2
  \overrightarrow{ \frac\partial {
      \partial\bar\zeta } }\; +\;
\overleftarrow{\frac{\partial}{ \partial\zeta }}
  \theta(|\zeta |^2)^2\;\overleftarrow{\frac{\partial}{ \partial\zeta }}
   \;\;
  \overrightarrow{ \frac{\partial^2} {
      \partial\bar\zeta^2 } }\biggr]\;\;+\;\;\cdot\cdot\cdot
\label{afsvt}\ee
where  $\theta(|\zeta|^2)$ is the covariant symbol of $
\tilde\Theta({\bf n})  $,
\be \theta(|\zeta|^2) = <\zeta|\tilde\Theta({\bf n})|\zeta> \ee
  Although we don't have an explicit
expression for $\theta(x)$ for our case, we can determine some of its
features.  At the
origin:
\be\theta(0) = <n_0|\tilde\Theta({\bf n})|n_0 > =\tilde\Theta(n_0)= \frac
{\Theta_0}{n_0+1}\;, \label{tnaogn}\ee  since $ |\zeta = 0>=|n_0>$.
From (\ref{tfln}), we also know that  $\theta(x)$
approaches a constant $\Theta_0$   as $x\rightarrow \infty$, and it is
not difficult to determine how fast it approaches this constant.  Because $\theta(x)$ varies slowly in this
region
we may approximate the star product with a finite number of terms. 
Now take the expectation value of (\ref{qdtz}) with respect to
coherent state $|\zeta >$:
\be <\zeta|\tilde\Theta({\bf n})^{-1}|\zeta> +
<\zeta|[\tilde\Theta({\bf n})^{-1},\bar z] z \tilde\Theta({\bf
  n})^{-1}|\zeta> + <\zeta|\bar z\tilde\Theta({\bf n})^{-1}[z, \tilde\Theta({\bf
  n})^{-1}]|\zeta>=\Theta^{-1}_0 \ee  To lowest order in the star
product this gives
\be   \theta(|\zeta|^2)^{-1} + \zeta\frac\partial {
      \partial\zeta }   \theta(|\zeta|^2)^{-1}
+\bar \zeta \frac\partial {
      \partial\bar \zeta } \theta(|\zeta|^2)^{-1}=\Theta^{-1}_0 \ee 
or
\be   \theta(x)^{-1} +2 x\frac d {d x} \theta(x)^{-1}  = \Theta^{-1}_0
\ee and so for large $x$\be \theta(x)^{-1} \approx  \Theta^{-1}_0\biggl(1
+\frac C{\sqrt{x}}\;\biggr)\;, \label{affth}\ee 
where $C$ is the same constant appearing in (\ref{tfln}).  For this we
used $<\zeta|{\bf n}|\zeta>\approx \Theta_0^2 |\zeta|^2$, which
is valid at
leading order.

We can also determine the asymptotic behavior of ${\cal N}(x)$  as
$x\rightarrow \infty$.  At lowest order we must recover the result for
the noncommuting plane, namely  ${\cal
  N}(x)\approx\exp{\Theta_0^{-1}x}$.  To get the next order we  make use of (\ref{nas}) and
(\ref{ddxln}) to write \be <\zeta |({\bf n}+1)\Sigma({\bf n})^{-1} |\zeta >
= \biggl(\frac d{dx}\ln{{\cal N}(x)} +\frac{n_0}x(1-{\cal
  N}(x)^{-1}\;)\biggr)\bigg|_{x=|\zeta |^2  } \ee  From (\ref{ndneqfmo}) the right hand
side is the covariant symbol of $\Theta_0 \tilde\Theta^{-2}({\bf n})$.  At
lowest order in the derivative expansion, we can then approximate the right hand
side by $\Theta_0 \theta(|\zeta |^2)^{-2}$, while on the left hand side we can drop
terms that go like $1/x$, so \be \Theta_0 \theta(x)^{-2}\approx 
\frac d{dx}\ln{{\cal N}(x)} \ee  Substituting the asymptotic solution
for $\theta(x)$ then gives  \be {\cal
  N}(x)\approx\exp{\Theta_0^{-1}x\biggl(1+ \frac{4C}{\sqrt x}}\biggr)\label{lxN}\ee

\section{The commutative limit}
\setcounter{equation}{0}

We next take the commutative limit in a way that preserves the
differential structure.
We define it as follows.  First re-scale
the variables $\zeta$ and $\bar\zeta$ by a factor $\frac1{\sqrt{\hbar}}$, and
then take the limit   $\hbar\rightarrow 0$.  Consequently, the
asymptotic behavior of functions such as  $\theta(x)$  also
corresponds to the
commutative limit.  Moreover, the order of the
derivative expansion of the star product
in (\ref{afsvt}) agrees with the order of $\sqrt{\hbar}$, and for
small
$\hbar$ we may approximate the star product with a finite number of
terms.
As required, at lowest order in $\hbar$
\beqa  [{\cal A}\star {\cal B}](\zeta,\bar \zeta )&\rightarrow &{\cal
  A}(\zeta,\bar \zeta )\;{\cal B}(\zeta,\bar \zeta )\cr
& &\cr[{\cal A}\star
{\cal B}-{\cal B}\star {\cal A}](\zeta,\bar \zeta )&\rightarrow &i\hbar\;
\{{\cal A},{\cal B}\}(\zeta,\bar \zeta )\;,\eeqa where the Poisson
bracket is defined by $$
\{{\cal A},{\cal B}\}(\zeta,\bar \zeta ) = -i\theta_c(\sqrt{2}|\zeta|) \;\biggl(
 \frac\partial{ \partial\zeta }{\cal
  A}(\zeta,\bar \zeta )\;\; 
 {\partial\over{ \partial\bar\zeta} }{\cal B}(\zeta,\bar \zeta
 )- \frac\partial{ \partial\zeta }{\cal
  B}(\zeta,\bar \zeta )\;\; 
 {\partial\over{ \partial\bar\zeta} }{\cal A}(\zeta,\bar \zeta
 )
\biggr)\;, $$ and where
$\theta_c(\sqrt{2}|\zeta|)\equiv \lim_{\hbar \rightarrow 0}
\theta(|\zeta|^2/\hbar)$.  If we introduce Cartesian coordinates
$$x^1=\frac1{\sqrt{2}}(\zeta +\bar\zeta)\;,\qquad
x^2=\frac1{\sqrt{2}i}(\zeta -\bar\zeta)\;,$$ then  $$
\{{\cal A},{\cal B}\}(\vec{x}) = \theta_c(r) \;\epsilon^{ij} 
\partial_i{\cal
  A}(\vec{x} )\; 
 \partial_j{\cal B}(\vec{x})
 $$ Here $\epsilon^{01}=-\epsilon^{10}=1$, $r=\sqrt{x^ix^i}$ and
 $\partial_i =\frac{\partial}{ \partial x^i}$, and the fundamental
 Poisson brackets are 
\be\{x^i ,x^j\}= \theta_c(r) \epsilon^{ij}\label{fpb}\ee

We can define the exterior
derivative of any function ${\cal A}(\vec{x})$ in a manner analogous
to (\ref{daqa}) in the full noncommutative theory, i.e. by taking the Poisson
bracket of  ${\cal A}(\vec{x})$ with the analogue of the operator $Q$,
call it now ${\cal Q}$,\footnote{This is similar to the approach
  followed in \cite{morar}.} \be d {\cal A}(\vec{x})= \{ {\cal Q},{\cal
  A}\}(\vec{x})\ee
In analogy to (\ref{odq}) we define
\be {\cal Q}=\theta_c(r)^{-1} \epsilon_{ij} x^i c^j\label{clsq}\;, \ee where
$c^i,\;i=1,2$  are a
basis of one forms
 and are the analogues of       $\chi$ and 
$\bar \chi$.  We assume $c^i$ have zero Poisson bracket with $x^i$ and
with themselves. 
 In general,  we cannot identify the basis of one
forms $c^i$ with $dx^i$ since 
\be dx^i\equiv \{Q,x^i\}= c^i 
+\epsilon_{ij}\epsilon_{k\ell}   \theta_c^{-1}\partial_j\theta_c x^k c^{\ell}\label{dxitc} \ee 
    For $d^2=0$, we need that $\{ {\cal Q},{\cal
  Q}\}(\vec{x})$ is in the center of the Poisson algebra.   If we define directional
derivatives ${\cal D}_i$, such that \be d {\cal A}(\vec{x})=[ {\cal D}_i {\cal
  A}](\vec{x})\;c^i\ee then this condition means that  $[{\cal D}_1, {\cal  D}_2]=0 $.
  From
(\ref{clsq}) we get  $$\{{\cal Q},{\cal Q}\}=(\theta_c^{-1} + x^k \partial_k
\theta_c^{-1})\;\epsilon_{ij}c^ic^j$$   This is in
the center of the Poisson algebra for  \be \theta_c^{-1} + x^k \partial_k
\theta_c^{-1}=\theta^{-1}_0 = {\rm constant}\label{dtz}\ee  The rotationally invariant solution  to (\ref{dtz}) is
\be\theta_c(r) =\frac{ \theta_0}{1 + \frac{r_0}r}\;, \label{ris} \ee where
 $r_0$ is  a constant.  We thus obtain a non constant Poisson
 structure.  In comparing with the asymptotic form of
 $\theta(|\zeta|^2)$ in the full noncommutative  theory
 [c.f. (\ref{affth})], $r_0$ and $\theta_0$ get identified with
 $\sqrt{2\hbar}\; C$ and $\Theta_0$, respectively.  The
 noncommutative plane at lowest order is
recovered when $r_0=0$.  For $r_0\ne 0$, $\theta_c(r)$ vanishes at the
origin, unlike  $\theta(|\zeta|^2)$ in the full noncommutative  theory
[c.f. (\ref{tnaogn})].   From (\ref{tnaogn}), we see that to get the
classical limit we must take $n_0\rightarrow\infty$, {\it in addition
  to} $\hbar\rightarrow 0$.  Furthermore, since differentiation involves
a factor $\theta_c(r)^{-1}$, we must remove the point at
the origin from ${\mathbb{R}}^2$.  The commutative limit is therefore a punctured
plane, which we denote by ${\cal M}$.
[If $r_0<0 $ there is, in addition, a singularity in $\theta$ (or a zero
in the density) at $-r_0$.  In that
case we can distinguish  the two regions $r<-r_0$ and $r>-r_0$.  The 
former corresponds to a disc with the central origin removed, and the
latter is ${\mathbb{R}}^2$ with a hole.]

Next we examine the orbits generated by the derivatives  ${\cal D}_i$
on ${\cal M}$, and determine an invariant measure.  These orbits are
the commutative limit of those obtained with the action of the unitary
operator $U(\beta,\bar\beta)$ in
sec. 4.  Infinitesimal motion  along the orbit is given by
$\delta_\epsilon x^i= \epsilon^j  {\cal D}_jx^i= \epsilon^j
\{\theta_c^{-1}\epsilon_{ji}x^j,{\cal A}\}$, $\epsilon^i$ being
infinitesimal.  From (\ref{clsq}) and (\ref{ris})
we get \be\delta_\epsilon x^i = \epsilon^i - \frac{r_0}{r+r_0}
\;(\epsilon^i - \epsilon\cdot \hat x\; \hat x^i )\;,\qquad \hat x = \vec{x}/r
\ee  The resulting orbits approach a constant direction as
$r\rightarrow \infty$, and point radially at the origin.  It is easy
to write an invariant measure $d\mu_I(\vec{x})$ with respect to these orbits.  Since  
$ {\cal D}_jc^i=0$, it is just $c^1c^2$.  Using (\ref{dxitc}) we can
express $d\mu_I(\vec{x})$ in terms of $x^i$ and $dx^i$.  The  Jacobian of the
transformation $c^i\rightarrow dx^i$ is just
${\theta_0}\theta_c^{-1}$, so
\be d\mu_I(\vec{x})= {\theta_0}\theta_c(r)^{-1}\; dx^1 dx^2 \label{inme}\ee 
Since $\theta_c(r)^{-1}$ appears in the invariant measure we can again argue
that the singular point at the origin must be removed from the manifold.

From the invariant measure we can associate a metric $ds^2$ on ${\cal M}$.
From the rotational symmetry
\be ds^2 = \gamma(r)\; dx^idx^i = 2\gamma(r)\; d\zeta d\bar\zeta\;, \ee
 with corresponding infinitesimal area element equal to $ \gamma(r)\; dx^1
 dx^2 $.  We get the metric upon identifying the area element with the invariant measure (\ref{inme}),
\be \gamma =  {\theta_0}\theta_c(r)^{-1} = 1+\frac{r_0}r \ee
Computing the scalar curvature we find
\be R= \frac{\frac{r_0}r(2+  \frac{r_0}r)}{(1+ \frac{r_0}r)^3}  \ee
Since $R$ is well behaved at the origin - it is only a coordinate singularity. 

  From (\ref{fpb}), we
see that the infinitesimal area element  $ \gamma(r)\; dx^1
 dx^2 $ (up to a scale factor) is the symplectic
two form for the theory.  Then by definition ${\cal M}$ is a two
dimensional homogeneous K\"ahler manifold.  The K\"ahler potential $\Gamma$ is
defined by \be \gamma = \frac\partial{ \partial\zeta }\frac\partial{
  \partial\bar \zeta }\; \Gamma\ee  for which we find \be 
\Gamma = |\zeta|^2 +2\sqrt{2} r_0 |\zeta| \label{kapt}\ee  From (\ref{lxN}), and
the identification of $r_0$  with
 $\sqrt{2\hbar}\; C$, we note that the classical limit  ${\cal
  N}_c$ of the normalization factor $ {\cal
  N}$ is related to $\Gamma$ by \be {\cal
  N}_c(\sqrt{2}|\zeta| ) = \lim_{\hbar\rightarrow 0}  {\cal
  N}(|\zeta |^2/\hbar ) = \exp{(\Gamma/{\tt h})} \;,\ee
where we define ${\tt h}=\hbar\Theta_0\;$.   This relation in fact shows up
for Berezin quantization on K\"ahler manifolds with a bounded domain
in ${\mathbb{C}}^n$.\cite{brzn},\cite{prlmf}   In comparing  Berezin's techniques to the ones we
employed\cite{us}, we remark that his expression for the star product
is formally the same as  ours, i.e. (\ref{gsp}).  To define it one only needs the scalar
product and the integration measure.  In  Berezin quantization,  the
scalar product can be expressed in terms of  normalization function $ {\cal
  N}$ in the same way as in our
approach (\ref{zetet}).   Unlike us, Berezin gives an explicit
 expression (\ref{eebm}) for the integration measure
$d\mu_B(\zeta,\bar\zeta)$, and it is written in terms of  `classical'
functions $\theta_c$ and $\Gamma$, as well as ${\cal N}$.  In Appendix B,
we show that this  integration measure $d\mu_B(\zeta,\bar\zeta)$
is not applicable to our
coherent states $|\zeta>$, as it leads to a violation of the partition
of unity.  So although Berezin's approach and ours coincide at lowest
order in ${\tt h}$ (or $\hbar$), they  
are nevertheless distinct quantization procedures. More
  precisely, the procedure of constructing coherent states $|\zeta>$
  and taking the commutative limit, is not
  the inverse of Berezin's quantization.  Therefore applying Berezin's
  quantization to
${\cal M}$  will lead to a different theory from the one
 we started with in sec. 3, and in all likelihood it will not have the
 feature that it admits an exterior
 derivative.

\bigskip
\noindent
{\bf Acknowledgment}

\noindent
We thank A.P. Balachandran, G. Karatheodoris and    P. Pre\u snajder
for valuable discussions.
This work was supported by the joint NSF-CONACyT grant E120.0462/2000 and
 the U.S. Department of Energy
 under contract number DE-FG05-84ER40141.

\bigskip
\noindent
\appendice

Here we prove (\ref{afsvt}) using the generalized coherent states
(\ref{ndcs}).
Following \cite{us}, given the scalar product $ <\zeta |\eta >$ and
the integration measure $ d\mu (\zeta,\bar\zeta )  $ satisfying
 the partition of unity\be 1=\int  d\mu (\zeta,\bar\zeta )  |\zeta><\zeta|\;,\label{aone}\ee the following closed form expression for the star
product between two functions
${\cal A}(\zeta,\bar \zeta )$ and ${\cal B}(\zeta,\bar \zeta )$ can be
given:
\be    [{\cal A}  \star {\cal B}](\zeta,\bar \zeta ) \;=\;  {\cal A}(\zeta,\bar \zeta )
 \;\;\int d\mu (\eta,\bar\eta )\;   \;
:\exp{ \overleftarrow{ \frac\partial{ \partial\zeta }}(\eta-\zeta)  }:
\;\;| <\zeta |\eta >|^2\;    \;
:\exp{  (\bar\eta-\bar\zeta ) \overrightarrow{ \frac\partial { \partial\bar\zeta } } }:\;\; {\cal B}(\zeta,\bar \zeta )
\label{gsp}\ee
The colons denote an ordered exponential such that the derivatives
don't act on the functions appearing in the exponent, i.e.
\beqa  :\exp{  (\bar\eta-\bar\zeta ) \overrightarrow{ \frac\partial {
       \partial\bar\zeta } } }: \;&=& 1 + (\bar\eta-\bar\zeta ) \overrightarrow{ \frac\partial {
       \partial\bar\zeta } } +\frac 12  (\bar\eta-\bar\zeta )^2 \overrightarrow{ \frac{\partial^2} {
       \partial\bar\zeta^2 } } +\cdot\cdot\cdot
\cr
:\exp{ \overleftarrow{ \frac\partial{ \partial\zeta }}(\eta-\zeta)
}: \;&=&1 + \overleftarrow{ \frac\partial{ \partial\zeta }}(\eta-\zeta)
+\frac12  \overleftarrow{ \frac{\partial^2}{ \partial\zeta^2 }}(\eta-\zeta)^2
 +\cdot\cdot\cdot\label{eooe} \eeqa   
The scalar product can be expressed in terms of ${\cal N}$ as
\be<\eta | \zeta >  = {\cal  N} (|\eta |^2  )^{-\frac12} \;{\cal N} (|\zeta |^2  )^{-\frac12}
{\cal  N}(\bar\eta\zeta) \;,  \label{zetet} \ee  and the measure $d\mu (\eta,\bar\eta )\;$ can be
written in terms of an inverse Mellin integral transform\cite{us}, but neither will  be
needed here. 
 Using the partition of unity
 (\ref{aone}), it is easy to see that at lowest order in the
 derivative expansion 
(\ref{eooe}),  $[{\cal A}  \star {\cal B}](\zeta,\bar \zeta
)$  reduces to the pointwise product \be       {\cal A}(\zeta,\bar \zeta )
 \;\;\int d\mu (\eta,\bar\eta )\;
\;\; <\zeta |\eta ><\eta |\zeta >\; {\cal B}(\zeta,\bar \zeta )\;
=\;{\cal A} (\zeta,\bar \zeta
) {\cal B}(\zeta,\bar \zeta
)\;,\ee  while both first order terms vanish:
\beqa  & &     {\cal A}(\zeta,\bar \zeta )
 \;\;\int d\mu (\eta,\bar\eta )\;   \;
 \overleftarrow{ \frac\partial{ \partial\zeta }}(\eta-\zeta)  
\;\; <\zeta |\eta ><\eta |\zeta >\;    \;
\; {\cal B}(\zeta,\bar \zeta ) \cr
&= &    \frac\partial{ \partial\zeta } {\cal A}(\zeta,\bar \zeta )
 \;\;\int d\mu (\eta,\bar\eta )\;   \;   
( <\zeta|z |\eta ><\eta |\zeta >-<\zeta |\eta ><\eta |z|\zeta > )   \;
\; {\cal B}(\zeta,\bar \zeta )\cr
&= &    \frac\partial{ \partial\zeta } {\cal A}(\zeta,\bar \zeta )
 \; \;   
( <\zeta|z |\zeta >-<\zeta |z|\zeta > )   \;
\; {\cal B}(\zeta,\bar \zeta ) \;=\;0
\eeqa
\beqa
  & &  {\cal A}(\zeta,\bar \zeta )
 \;\;\int d\mu (\eta,\bar\eta )\;   \;
 <\zeta |\eta ><\eta |\zeta >\;    \;
  (\bar\eta-\bar\zeta ) \overrightarrow{ \frac\partial { \partial\bar\zeta } } \;\; {\cal B}(\zeta,\bar \zeta )\cr
 &= &  {\cal A}(\zeta,\bar \zeta )
 \;\;\int d\mu (\eta,\bar\eta )\;   \;
 (<\zeta |\eta ><\eta|\bar z |\zeta >- <\zeta|\bar z |\eta ><\eta |\zeta >)    \;
   \frac\partial {
      \partial\bar\zeta }  {\cal B}(\zeta,\bar \zeta )\cr
 &= &  {\cal A}(\zeta,\bar \zeta )
 \;   \;
 (<\zeta |\bar z |\zeta >- <\zeta|\bar z |\zeta >)    \;
   \frac\partial {
      \partial\bar\zeta }  {\cal B}(\zeta,\bar \zeta )\;=\;0 \eeqa
The only nonvanishing second order term is
\beqa  & &     {\cal A}(\zeta,\bar \zeta )
 \;\;\int d\mu (\eta,\bar\eta )\;   \;
 \overleftarrow{ \frac\partial{ \partial\zeta }}\;\;|\eta-\zeta |^2  
 <\zeta |\eta ><\eta |\zeta >\; 
  \overrightarrow{ \frac\partial { \partial\bar\zeta } } \;\; {\cal B}(\zeta,\bar \zeta )\cr
 &= & \frac\partial{ \partial\zeta } {\cal A}(\zeta,\bar \zeta )
 \;\;\int d\mu (\eta,\bar\eta )\;   \;
 (<\zeta |z|\eta ><\eta |\bar z|\zeta >+ <\zeta| \bar z |\eta ><\eta |z |\zeta
 >\cr &  & \qquad\qquad -<\zeta|\bar z z |\eta ><\eta |\zeta >-<\zeta |\eta ><\eta |
\bar z z|\zeta >)    \;
   \frac\partial {
      \partial\bar\zeta }  {\cal B}(\zeta,\bar \zeta )\cr
 &= & \frac\partial{ \partial\zeta } {\cal A}(\zeta,\bar \zeta )
 \;   \;
 (<\zeta |z\bar z |\zeta >- <\zeta|\bar zz |\zeta >)    \;
   \frac\partial {
      \partial\bar\zeta }  {\cal B}(\zeta,\bar \zeta )\cr
 &= & \frac\partial{ \partial\zeta } {\cal A}(\zeta,\bar \zeta )
 \;   \;
 \theta(|\zeta |^2)    \;\;
   \frac\partial {
      \partial\bar\zeta }  {\cal B}(\zeta,\bar \zeta )
 \eeqa
agreeing with  (\ref{afsvt}).  Applying the same procedure at the next
order gives the terms $$  \frac12   
 \;\;\int d\mu (\eta,\bar\eta )\;   \;
 \overleftarrow{ \frac{\partial^2}{ \partial\zeta^2 }}\;\;|\eta-\zeta
 |^2 (\eta -\zeta)  
 <\zeta |\eta ><\eta |\zeta >\; 
  \overrightarrow{ \frac{\partial} { \partial\bar\zeta } } \;\;+\;\;{\rm
      complex \;conjugate} $$
\be =
\frac12\;\overleftarrow{\frac{\partial^2}{ \partial\zeta^2 }}
    \;
 <\zeta|[z,\tilde\Theta({\bf n})]|\zeta>   \;
   \overrightarrow{\frac\partial {
      \partial\bar\zeta }}\; \;+\;\;{\rm
      complex \;conjugate}
\ee  acting on $ {\cal A}(\zeta,\bar \zeta )$ to the right
and $ {\cal B}(\zeta,\bar \zeta )$ to the left. Although
these terms involve three derivatives acting on ${\cal A}$ and ${\cal B}$, overall,
it is fourth order in derivatives since
\be <\zeta|[z,\tilde\Theta({\bf n})]|\zeta>  =\zeta \star \theta(|\zeta |^2)  
- \theta(|\zeta |^2)   \star \zeta\ee
which is $  \theta(|\zeta |^2) \frac\partial {
      \partial\bar\zeta }\theta(|\zeta |^2)$ to leading order.  The
    result is  \be
\frac12\;\overleftarrow{\frac{\partial^2}{ \partial\zeta^2 }}
    \; \theta(|\zeta |^2) \frac\partial {
      \partial\bar\zeta }\theta(|\zeta |^2)
   \;
  \overrightarrow{ \frac\partial {
      \partial\bar\zeta } }\; \;+\;\;{\rm
      complex \;conjugate}\label{fot1} \ee
Another fourth order term comes from 
$$  \frac14    
 \;\;\int d\mu (\eta,\bar\eta )\;   \;
 \overleftarrow{ \frac{\partial^2}{ \partial\zeta^2 }}\;\;|\eta-\zeta |^4  
 <\zeta |\eta ><\eta |\zeta >\; 
  \overrightarrow{ \frac{\partial^2} { \partial\bar\zeta^2 } } \;\;$$
\be  =\frac14\; \overleftarrow{\frac{\partial^2}{ \partial\zeta^2 }}
    \;
 <\zeta|2\tilde\Theta({\bf n})^2 +[\bar z,\tilde\Theta({\bf
   n})]z+z[\tilde\Theta({\bf n}),\bar z]
|\zeta>   \;
   \overrightarrow{ \frac{\partial^2} {
      \partial\bar\zeta^2 } } \ee  from
    which we extract the lowest order contribution:
\be\frac12 \;\overleftarrow{\frac{\partial^2}{ \partial\zeta^2 }}
    \; \theta(|\zeta |^2)^2   \;
    \overrightarrow{\frac{\partial^2} {
      \partial\bar\zeta^2 }}\label{fot2}\ee
The fourth order terms (\ref{fot1}) and (\ref{fot2}) can be combined
to give the result in (\ref{afsvt}).

\bigskip
\noindent
\appendice

Here 
we show that Berezin's  integration measure $d\mu_B(\zeta,\bar\zeta)$
 violates the partition of unity (\ref{aone}) for our
coherent states $|\zeta>$.  Our proof is by contradiction.  $d\mu_B(\zeta,\bar\zeta)$ is defined to be\cite{brzn},\cite{prlmf}
\beqa
d\mu_B(\zeta,\bar\zeta)& =& i
H (|\zeta|^2  ) \;d\zeta \wedge d\bar\zeta\;,\cr & &\cr
  H (|\zeta|^2  )& =&-\frac{\alpha({\tt h})}{2\pi
}\theta_c(\sqrt{2}|\zeta|)^{-1}\exp\left(-\Gamma\right /{\tt h}){\cal N}(|\zeta|^2)\ ,\label{eebm}\eeqa
where $\alpha({\tt h})$ is some normalization constant.   For our system
$\theta_c$ and $\Gamma$ are given by (\ref{ris})  and (\ref{kapt}),
respectively, and ${\tt h}=\hbar\Theta_0\;$.
 Using (\ref{zetet})
the partition of unity (\ref{aone}) can be written  \be {\cal N}(\bar \zeta \lambda )=
\int d\mu (\eta,\bar\eta )\;   \;
 \frac {{\cal N}(\bar\zeta\eta) \;{\cal   N}(\bar\eta\lambda)  }
  { {\cal N} (|\eta |^2  )   }  \;, \label{zetlam} \ee  for arbitrary complex coordinates $\zeta$
and $\lambda$.  After substituting (\ref{eebm}), the integration over
the angular variable
can be easily performed, leaving 
\be I(n)\equiv \int_0^\infty \frac{d\rho \;\rho^{2n+1}  \; H (\rho^2  )}{ {\cal
    N} (\rho^2  )  }
 = \frac{1}{4\pi}\left\{\matrix{ 1\;, & n=0
      \cr\hbar^n\prod_{m=n_0}^{n-1+n_0}\Sigma(m)\;, &
 n\ge 1}\right.\label{coh}\;,\ee where factors of $\hbar$ came
from the rescaling of the coordinates.  Substituting in the expression for $H(\rho^2)$ from (\ref{eebm}), the
integral can be expressed in terms of parabolic-cylinder functions $D_n(u)$\cite{whit}
\be
I(n)=-\frac{\alpha({\tt  h})e^\frac{r_0^2}2\;(2n)!}{2\pi\Theta_0}\;\biggl(\frac {\tt h}2\biggr)^{n+1}
\;\biggl((2n+1)\;D_{-2n-2}(u)
+\frac u2 \;D_{-2n-1}(u)\biggr) \;,
\ee  where $u={2r_0}/{\sqrt{{\tt h}}}\;.$
  (\ref{coh}) for  $n=0$ can be used to fix the normalization
  constant $\alpha({\tt h})$, yielding
\be
\alpha({\tt h})=-\frac{\;e^{-\frac{r_0^2}2}\;\;\Theta_0/{\tt
    h}}{D_{-2}(u)
+\frac u2 \;D_{-1}(u)}\;,
\ee
 Substituting this result for $\alpha({\tt h})$    in (\ref{coh}) for  $n\ge1$ gives
\be  (2n)!\;\biggl(\frac {\Theta_0}
 2\biggr)^{n}\;\frac{(2n+1)\;D_{-2n-2}(u)
+\frac u2 \;D_{-2n-1}(u)}{D_{-2}(u)
+\frac u2  \;D_{-1}(u)}\;=\;\prod_{m=n_0}^{n-1+n_0}\Sigma(m)
\label{bsix}\ee
This equality works for the case of the noncommutative plane
${r}_0=u=0$.  In this limit, $D_{-2n-2}(0) =2^n n!/(2n+1)!\;,$ and both the left and right hand sides reduce to $n!\;
\Theta_0^n$.  For the case of nonzero $r_0$, we can use one of the
conditions (\ref{bsix}) to determine the  relationship between ${r}_0$
and $n_0$.  Take for example $n=1$.  Then using (\ref{icft}),
\be  \frac{3\;D_{-4}(u)
+\frac u2 \;D_{-3}(u)}{D_{-2}(u)
+\frac u2  \;D_{-1}(u)}\;=\frac 1{n_0+1}
\label{bsev}\ee   
Solving this in general and substituting back into (\ref{bsix}) for
 $n\ge 2$ is
 nontrivial.  However (\ref{bsix}) should hold for all $n_0$, and to
 disprove it only requires a counter example.  For this purpose we look at the case of large $n_0$.  This means we are approaching  the
 commutative limit  where $u$ must also be large.  Here we can use the
following asymptotic expansion for the  parabolic-cylinder functions
\be
D_p(u) = e^{-\frac{u^2}{4}}u^p(1+O(u^2))
\ee
At leading order we get $\;\;u\rightarrow  \sqrt{n_0}\;\;$ or
 $\;\;r_0\rightarrow 
\frac 12 \sqrt{\hbar n_0 \Theta_0}\;.$  Substituting this result into
 the left hand side of (\ref{bsix}) for $n=2$ gives $\;\;6 (\Theta_0/n_0)^2\;\;$ to
 leading order.  Using  (\ref{icft}) and (\ref{nzpo}),  the right hand
 side  of (\ref{bsix}) for $n=2$ instead gives $\;\; (\Theta_0/n_0)^2
 (\sqrt{5}+1)/(\sqrt{5}-1)\;.$  So the relation (\ref{bsix})   fails and
 Berezin's  integration measure $d\mu_B(\zeta,\bar\zeta)$ is not ours.

\end{document}